\documentclass[12pt,preprint]{aastex}

\shorttitle{Chasing shadows in the disk of TW Hya} 
\shortauthors{Debes et al.}
\begin{document}
\title{Chasing Shadows: Rotation of the Azimuthal Asymmetry in the TW Hya Disk\footnote{Based on observations made with the NASA/ESA Hubble Space Telescope, obtained from the data archive at the Space Telescope Science Institute. STScI is operated by the Association of Universities for Research in Astronomy, Inc. under NASA contract NAS 5-26555.}}
\author{John H. Debes\altaffilmark{1}, Charles A. Poteet\altaffilmark{1}, Hannah Jang-Condell\altaffilmark{2}, Andras Gaspar\altaffilmark{3}, Dean Hines\altaffilmark{2}, Joel H. Kastner\altaffilmark{4}, Laurent Pueyo\altaffilmark{1}, Valerie Rapson\altaffilmark{4}, Aki Roberge\altaffilmark{5}, Glenn Schneider\altaffilmark{3},Alycia J. Weinberger\altaffilmark{6}}
\altaffiltext{1}{Space Telescope Science institute, Baltimore, MD 21218, U.S.A.}
\altaffiltext{2}{Department of Physics \& Astronomy,  University of Wyoming, Laramie, WY 82071, U.S.A.}
\altaffiltext{3}{Steward Observatory, The University of Arizona, Tucson, AZ 85721, U.S.A.}
\altaffiltext{4}{Department of Astronomy, Rochester Institute of Technology, Rochester, NY 14623, U.S.A.}
\altaffiltext{5}{NASA Goddard Spaceflight Center, Greenbelt, MD 20771, U.S.A.}
\altaffiltext{6}{Department of Terrestrial Magnetism, Carnegie Institution for Science, Washington, DC 20015,U.S.A.}

\begin{abstract}
 We have obtained new images of the protoplanetary disk orbiting TW Hya in visible, total intensity light with the Space Telescope Imaging Spectrograph (STIS) on the Hubble Space Telescope (HST), using the newly commissioned BAR5 occulter. These HST/STIS observations achieved an inner working angle $\sim$0.2\arcsec, or 11.7~AU, probing the system at angular radii coincident with recent images of the disk obtained by ALMA and in polarized intensity near-infrared light. By comparing our new STIS images to those taken with STIS in 2000 and with NICMOS in 1998, 2004, and 2005, we demonstrate that TW Hya's azimuthal surface brightness asymmetry moves coherently in position angle.  Between 50~AU and 141~AU we measure a constant angular velocity in the azimuthal brightness asymmetry of 22.7$^\circ$~yr$^{-1}$ in a counter-clockwise direction, equivalent to a period of 15.9~yr assuming circular motion. Both the (short) inferred period and lack of radial dependence of the moving shadow pattern are inconsistent with Keplerian rotation at these disk radii. We hypothesize that the asymmetry arises from the fact that the disk interior to 1~AU is inclined and precessing due to a planetary companion, thus partially shadowing the outer disk.  Further monitoring of this and other shadows on protoplanetary disks potentially opens a new avenue for indirectly observing the sites of planet formation.
  \end{abstract}


\section{Introduction}
TW Hya is a nearby protoplanetary disk-hosting star that has been heavily studied over a broad range of wavelengths \citep[e.g.,][]{brickhouse,france,kastner,nomura16}.  Its nearly pole-on disk \citep[i$<$7$^\circ$;][]{qi04} displays indications of active planet formation, despite its inferred age of 7-10~Myr \citep{weinberger13,herczeg14,ducourant14}.  Features in the disk that are indicative of such planet formation include an inner clearing in the disk out to $\sim$2.6~AU for mm-sized dust \citep{andrews16}, a partially cleared gap at 89~AU in scattered optical and Near-IR (NIR) light \citep{krist00,Weinberger:2002,debes13}, and a gap at 22-27~AU observed in polarized scattered light \citep{a15,rapson15}. A new interior gap has also recently been reported at 7~AU by \citet{vanboekel}. 

We note that TW Hya's distance as determined by Gaia, 59.5 pc \citep{tgas}, is roughly 10\% larger than, but within 1-$\sigma$ of, that determined from Hipparcos data \citet{vanLeeuwen:2007} using the Hipparcos mission. The distance for TW~Hya is still within 1-$\sigma$ of the Hipparcos value. We have thus adopted the {\em Gaia} value for the parallax, and all linear distances in AU reflect this slight change compared to previous discussions in the literature.

We have re-observed the TW Hya disk using the newly commissioned BAR5 occulter on the Hubble Space Telescope/Space Telescope Imaging Spectrograph (HST/STIS) to achieve an inner working angle (IWA) of 0\farcs2, which rivals that of groundbased polarized light observations.  The images push to smaller angular distances from the host star than previous STIS images of the disk taken at the WEDGEA1.0 position \citep{Roberge:2005}, enabling the exploration of the inner disk structure in scattered light. In addition to pushing inwards in visible light, these new STIS images allow us to investigate whether there are any time variable features in the disk's surface brightness distribution, as has been observed in other nearby  disks \citep[e.g., MP Mus,s][]{schneider14,wolff16}.

In \S \ref{sec:obs} we describe the observations of TW Hya with STIS, while in \S \ref{sec:analysis} we present an investigation of the inner surface brightness (SB) of the disk and report the apparent rotation of an azimuthal asymmetry present in the disk. In \S \ref{sec:physical} we determine some possible physical mechanisms that could account for the variability we observe, and in \S \ref{sec:conc} we discuss the implications of our observations.
  
\section{Observations}
\label{sec:obs}
Two new images of TW~Hya were obtained with the STIS CCD on 24-May-2015 and 24 March-2016 (1024$\times$1024 pixels, 50.7 mas/pixel), separated by roughly a year (GO \#13753; PI: Debes).  Table \ref{tab:obslist} lists all TW Hya observations used in this manuscript.  In each new STIS epoch, two contiguous orbits targeting TW Hya were executed, immediately followed by one orbit targeting a point spread function (PSF) reference, HD 88512, and finally an orbit of TW Hya. We used the same PSF star with earlier epochs of STIS data, to remove one source of uncertainty if temporal changes were observed. TW Hya was observed at three different sky orientations separated by $\pm$15$^\circ$ at each epoch.   

Within each orbit, the star was placed behind the BAR5\footnote{http://www.stsci.edu/hst/stis/strategies/pushing/coronagraphic\_bars} (width=0.15\arcsec) occulter.  A 3-point dither was executed with a width of $\pm$0.25 pixels ($\pm$13~mas) in the direction perpendicular to the long-axis of the occulter, and then the star was placed behind the WEDGEA1.0 (width=0.5\arcsec) occulting position.  Dithering mitigates contrast loss due to target acquisition non-repeatabilities. Exposures at BAR5 were short, to ensure the star did not exceed the CCD full well.  We utilized sub-arrays to minimize overheads due to readouts, with the BAR5 subarray consisting of a 1024x100 (51.9\arcsec$\times$5.071\arcsec) section of the detector centered on TW Hya and HD 85512.  The WEDGEA1.0 subarrays consisted of 1024x427 (51.93\arcsec$\times$21.65\arcsec) sections of the detector. In the first epoch, we obtained 54 exposures with integration times of 17~s each, and 9 exposures with 527s integration times at WEDGEA1.0.  Because the light in the PSF at the edge of the bar was smaller than originally estimated by roughly a factor of 5, we increased the BAR5 exposure times for the second epoch, such that there were now 18 exposures of 91~s each.  For reference, the average observed countrate per pixel at the edge of BAR5 on the unfiltered STIS CCD with GAIN=4 for TW Hya was 140 cts/s.  The WEDGEA1.0 exposure times were unchanged for the second epoch.  Total exposure times for each epoch are listed in Table \ref{tab:obslist}.

TW Hya was also observed over two epochs separated by two months in 2000 with STIS/WEDGEA1.0, also using HD~85512 \citep[GO \#8624; PI:Weinberger,][]{Roberge:2005}.  The spacecraft orientation difference between the two 2000 epochs was 33$^\circ$ for the science target.  In 2000, a background star near HD~85512 is present at $\sim$0.8\arcsec.  At one orientation it is near a diffraction spike, whereas at the second orientation it is close to the inner region of the disk.  By 2015-2016, the background star had moved $>$4\arcsec\ away due to HD 85512's high proper motion.  We isolated the star from HD~85512's PSF by subtracting one 2000 epoch from the other and creating an empirical PSF from the combined image.  We then subtracted this template from the original images.  We also masked other background stars that imprinted negative residuals onto our PSF subtracted images.  

We aligned HD~85512's position to that of TW~Hya and subtracted off a scaled reference PSF.  We centroided images by linearly fitting the vertical position of the diffraction spikes and solving for the intercept of the spikes to an accuracy of better than 0.1~pixel \citep{schneider14}.  Further refinements to the measured offsets and scalings were obtained by iteratively subtracting the scaled PSF from the target and minimizing a $\chi^2$ metric for centroid shifts and scalings.  For BAR5, we minimized subtraction residuals at the mask edge by selecting the best matches in dither position.  Finally, we applied a custom coronagraphic flatfield to all data, which was constructed from available flatfield images of the 50CORON aperture\footnote{Programs 12406, 13136,13539, and 13986}.  This corrects flux right near the occulters at the few percent level.  

We reverified the intensity scaling relations used between HD~85512 and TW~Hya in \citet{Roberge:2005}.  In the STIS passband, TW Hya has been variable at the level of 5-10\%.  The scalings used in PSF subtraction are presented in Table \ref{tab:obslist}.  We confirmed that TW Hya is not variable by greater than 2\% orbit-to-orbit via target vs. target subtractions in our 2015 and 2016 epochs.  We applied the stellar intensity scaling determined from the WEDGEA1.0 observations also to those at the BAR5 position where the disk fills the subarray.  For the 2015 and 2016 epochs, we combined the WEDGEA1.0 and BAR5 PSF-subtracted images by averaging overlapping areas near the occulters out to 2\arcsec.  Missing data at one position was filled with data from the other, if any existed.

STIS high contrast images can be particularly sensitive to color mismatches, which tend to impact disk photometry with azimuthally symmetric zones of over or under subtraction of the target PSF interior to $\sim$1\arcsec \citep{grady03,schneider14}.  HD~85512's spectral energy distribution does not exactly match TW~Hya across the STIS CCD bandpass, but the impact to PSF subtraction appears negligible at the level of $<$10\%, given the good agreement in the slope of the surface brightness at these distances compared to those measured in the Near-IR with NICMOS, GPI, and HiCIAO (See \S \ref{sec:analysis}).  

We noted that there existed a small difference in fractional disk brightness between BAR5 and WEDGEA1.0.  This is likely due to differing charge transfer efficiency at each occulter location from radiation damage on the STIS/CCD detector \citep{anderson10}.  The exact impact of these effects is dependent on the clocking of the detector, the temperature of the detector, the amount of charge within each pixel, the distance each row of the CCD is from the amplifier, and the background present.  Given the complexity of such an issue, we did not try to quantify these effects.  Mitigating this problem by scaling up the WEDGEA1.0 position images by 5\% allowed for a better empirical match between the BAR5 and WEDGEA1.0 images.  The effect adds an overall 5\% systematic uncertainty to the absolute SB of the disk at radii where BAR5 and WEDGEA1.0 overlap.

In addition to our STIS observations, we also used NICMOS images from programs 7226 (PI: Becklin), 10167 (Pi: Weinberger), and 9786 (PI: Hines).  The latter program in part consists of 2 $\micron$ coronagraphic imaging polarimetry of \object{TW Hya} from 2004.  Details describing the NICMOS polarimetric observations of \object{TW Hya}, as well as an in-depth analysis of the Stokes images, will be presented elsewhere (C.\ A.\ Poteet et al.\ 2016, in preparation).  In this study, however, we utilize the 2 $\micron$ polarimetric total intensity image (Stokes \emph{I}) of \object{TW Hya}, and briefly describe its construction.

Following the standard observing strategies for NICMOS coronagraphy, \object{TW Hya} was observed in February and April 2004, while and the unpolarized PSF reference star \object{BD$+$32$\degr$3739} was observed between November 2003 and August 2004 with the NICMOS Camera 2 linear polarizers (POL0L, POL120L, and POL240L).  The data were processed using high-level science products from the Legacy Archive PSF Library And Circumstellar Environments database \citep[LAPLACE;][]{Schneider10}, which are re-calibrated using improved techniques in flat-field correction, dark subtraction, and bad pixel correction.  During our processing procedure, the images were optimally registered using only the diffraction pattern produced by the \emph{HST} secondary mirror support.  In addition, the data were co-aligned in the science instrument aperture frame, background subtracted, and further examined for the presence of bad pixels.  Finally, the visit-specific co-aligned images were median-combined for each linear polarizer. 

To remove the PSF component from the \object{TW Hya} polarimetric data, classical PSF subtraction was implemented using the unpolarized star \object{BD$+$32$\degr$3739}.  After masking the central region surrounding the coronagraph, the scale factor between the \object{TW Hya} and \object{BD$+$32$\degr$3739} median-combined images was determined for each linear polarizer by minimizing the residual signal along the full extent of the diffraction pattern.  Adopting the polarimetry reduction algorithm of \citet{Hines00}, co-oriented Stokes \emph{I}, \emph{Q}, and \emph{U} images were produced from the visit-specific PSF-subtracted images, and subsequently smoothed using a three-pixel FWHM Gaussian.  The resulting images for all visits were then median-combined to construct the final Stokes images.   

\section{Multi-Epoch Imaging of the TW Hya Disk}
\label{sec:analysis}
\begin{figure}
\plotone{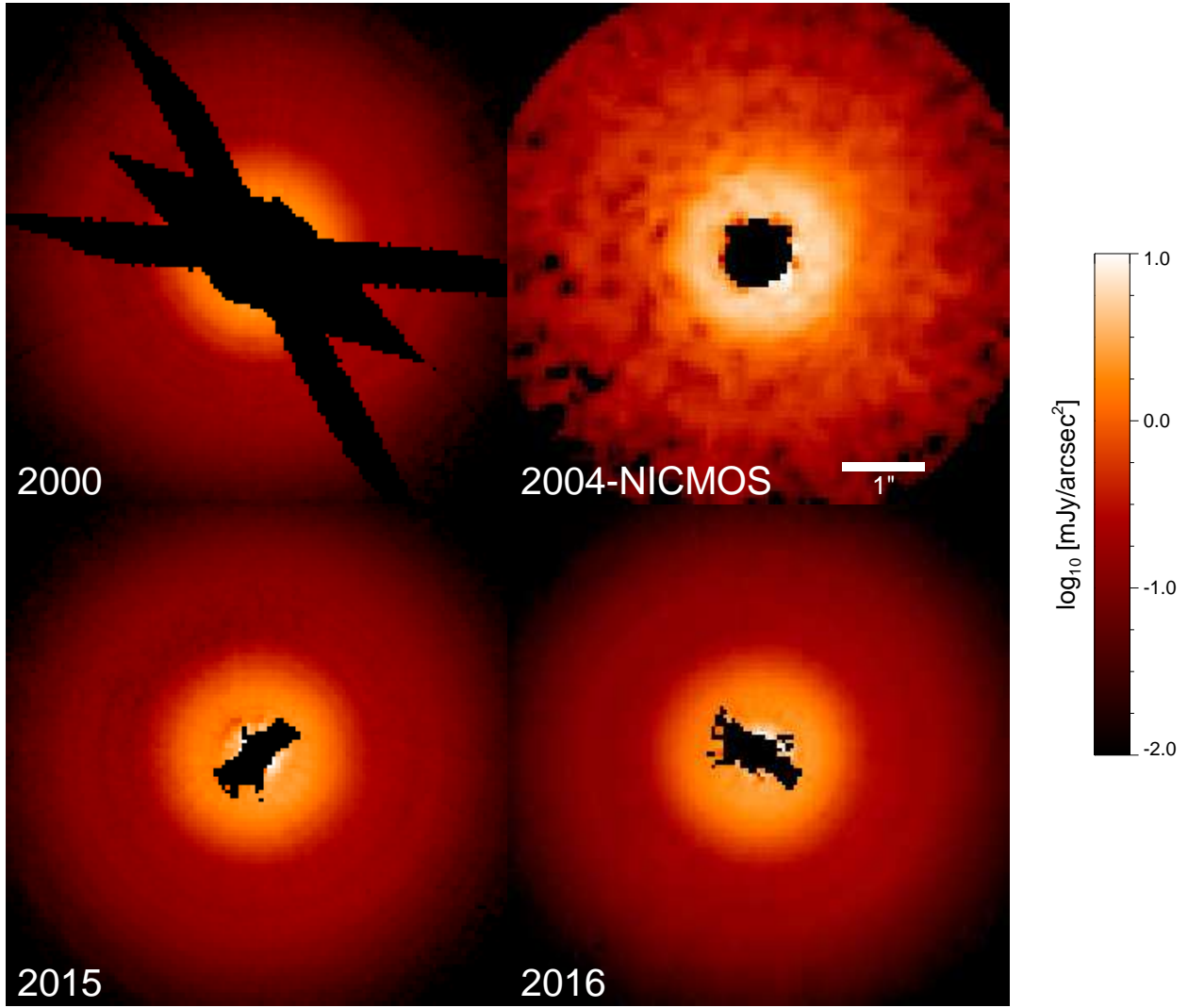}
\caption{\label{fig:images} Comparison of TW Hya over four epochs with STIS and NICMOS.  The three STIS images include the 2000 WEDGEA1.0 observations, as well as the 2015 and 2016 combination (WEDGEA1.0 and BAR5 occulter) observations.  Finally, the Stokes I (total intensity) image is from the 2004 NICMOS POL*L observations.  All images are logarithmically scaled, with North up and East left.  Missing data behind the coronagraphic occulters is displayed in black.}
\end{figure}

\begin{figure}
\plotone{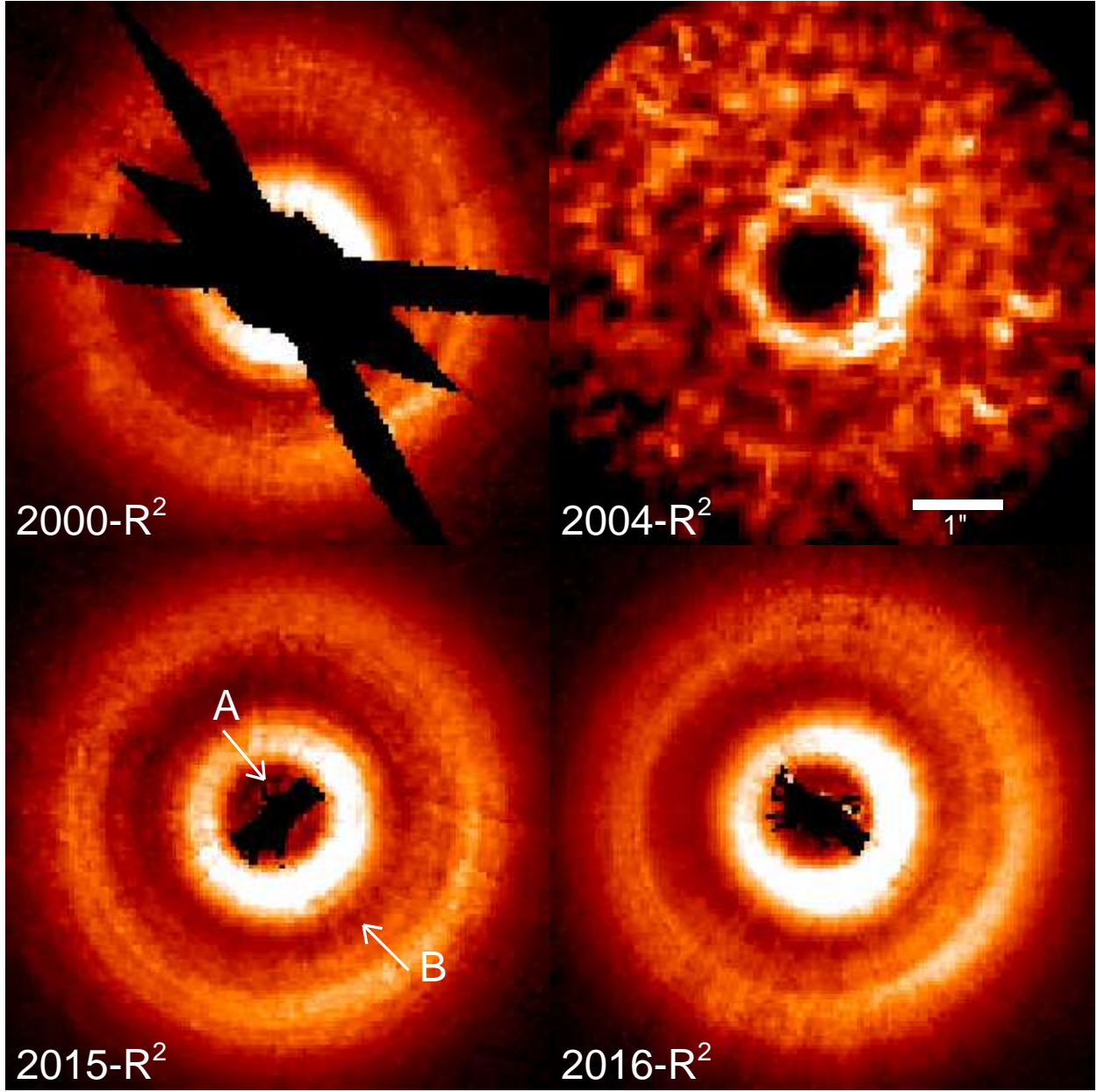}
\caption{\label{fig:imagesrsq} Same as in Figure \ref{fig:images}, but with each pixel scaled by its distance from the central star. Arrows A and B point to apparent gaps or shadows in the surface of the disk at 0\farcs4 and 1\farcs5 respectively.}
\end{figure}

Figure \ref{fig:images} shows logarithmically scaled images of the year 2000, 2004(NICMOS POL*), 2015, and 2016 epochs.  The latter two epochs have full angular coverage down to $\sim$0\farcs5 (29~AU) from the star, and partial angular coverage down to $\sim$0\farcs2 (11.8~AU).  For the STIS images we assumed a photometric conversion of 8.43e-7 Jy ct$^{-1}$~s~pixel$^{-1}$ at an effective wavelength of 7204\AA\, as given by a Pysynphot calculation \citep{pysynphot} of STIS' sensitivity for an M0 spectral type. We choose an M0 spectral type for TW~Hya in this wavelength range because this most closely represents the combination of the stellar photosphere and its veiling due to accretion \citep{debes13}.  In 2000 the photometric conversion is 1.6\% lower due to higher sensitivity of the instrument. We present images in Figure \ref{fig:imagesrsq} where the brightness within each pixel is scaled by the square of the stellocentric distance to highlight features within the disk. The inner part of the disk appears dim in these images because the disk SB profile is shallow from the inner working angle of the images out to 0\farcs7 (38~AU). Between 0\farcs7-1\arcsec\ (42-59~AU), a ring is present, with clear azimuthal asymmetries.  A second radial gap is clearly seen at 1.5\arcsec\ (89~AU), with a second fainter ring at 2\farcs3 (137~AU).

We measured azimuthally averaged SB profiles of TW Hya's disk from 2000, 2015, and 2016.  The uncertainty in our PSF scaling translates into at most a 5\% systematic uncertainty in disk SB per pixel.  Any residual systematic uncertainties, which dominate over statistical uncertainties, were accounted for by measuring the standard deviation in our photometric apertures.  We compare our SB profiles to those seen in polarized NIR light and to the ALMA dust continuum maps at high resolution.  In addition to our STIS observations, we also use NICMOS images from programs 7226 (PI: Becklin) and 10167 (PI: Weinberger).

\subsection{Radial surface brightness of TW Hya's disk}
\label{sec:SB}
\begin{figure}
\plottwo{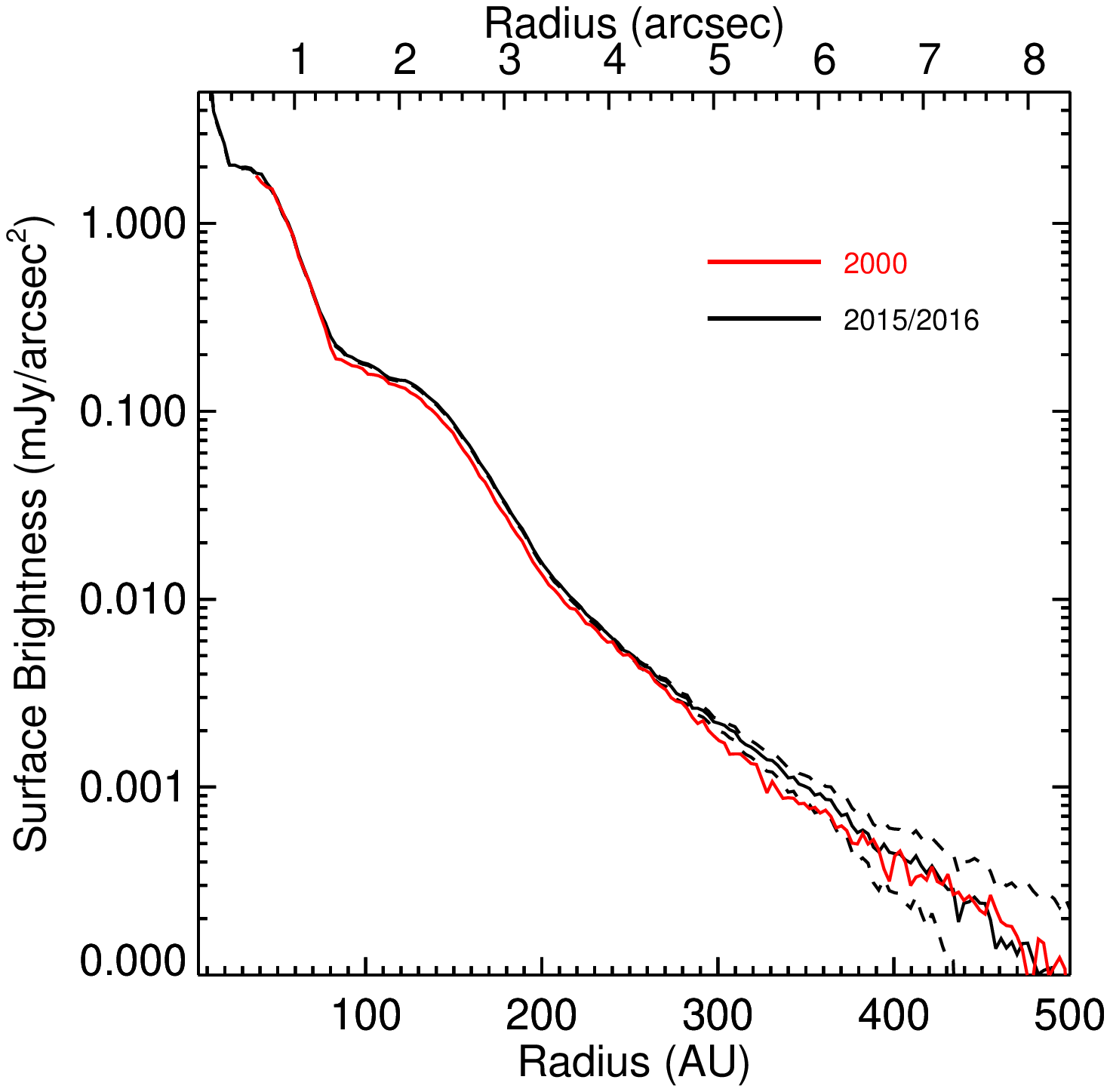}{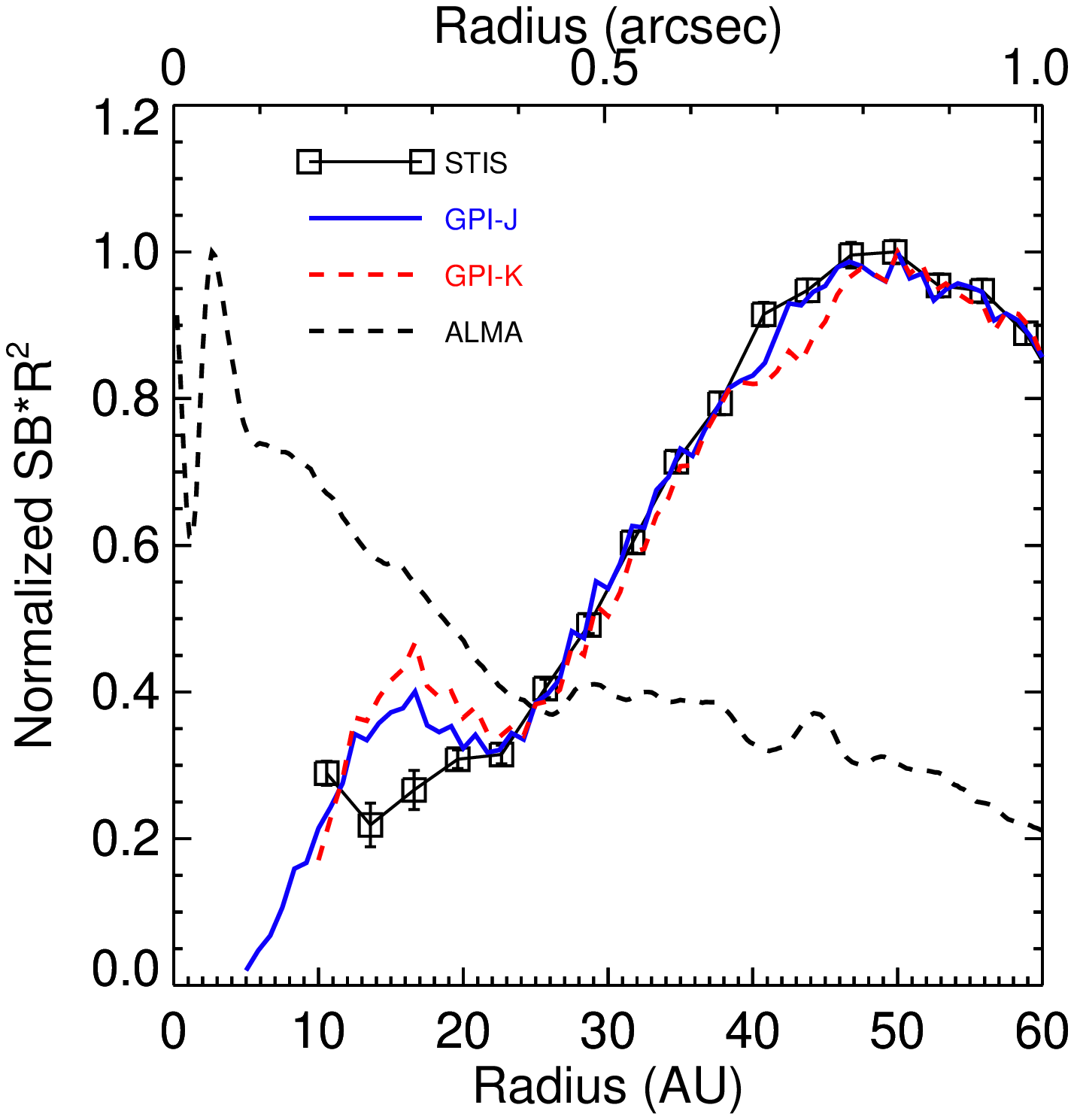}
\caption{\label{fig:analysis} (left) The TW Hya azimuthally averaged SB profile in 2015 (black line) and 2000 (red line) extending out to the detection limit of the disk. Photometric uncertainties on disk SB are equivalent to the thickness of the lines interior to 5\arcsec. The dashed black lines show the 1-$\sigma$ uncertainty in SB due to background subtraction, which becomes significant beyond 5\arcsec.  (right) Azimuthally averaged 2015+2016 SB (black squares) multiplied by R$^2$.  We overplot the GPI J and K band polarized intensity scaled profiles (blue and red lines) and the ALMA brightness temperature.  The ALMA data is normalized to 1 at its peak brightness, but not scaled by R$^2$.} 
\end{figure}

The left panel of Figure \ref{fig:analysis} shows the 360-degree azimuthally averaged SB profiles for the 2000 epoch compared to a combination of 2015/2016 taken with STIS.  We recover the broad behavior of the disk as seen in previous work \citep{a15,rapson15,debes16}. In particular, we confirm the presence of a depletion of SB from our inner working angle of 10~AU to 38~AU.  Following \citet{a15}, we report the STIS SB power-law fits to the region between 0.2-0.4\arcsec, and 0.4-0.8\arcsec.  We find slopes of -1.2$\pm$0.2 and -0.4$\pm$0.1, compared to -1.7 and -0.4 for the recent GPI images of the disk \citep{rapson15} and -1.4 and -0.3 for the recent HiCIAO image of the disk \citep{a15}.  The right panel of Figure \ref{fig:analysis} shows the excellent match between the total intensity light in the visible, and that of polarized intensity light at J and K.  

We averaged the images from the 2015 and 2016 epochs to determine the extent of the disk (left panel of Figure \ref{fig:analysis}).  The resulting azimuthally averaged profile shows disk flux at 10-$\sigma$ out to 7.6\arcsec, or 452~AU, which is larger than previously reported \citep{Roberge:2005}. Beyond this point, the disk is not detected significantly due to the systematic uncertainty in the background level of the images, which is denoted by the dashed lines in the left panel of Figure \ref{fig:analysis}.  Using our above flux conversions, the SB of the disk edge is 0.2$\mu$Jy arcsec$^{-2}$ at 7.6\arcsec. Comparing this limiting SB to the star, the per-pixel contrast of the disk relative to the flux density of the star \citep[0.18~Jy,][]{debes13} at this angular distance is 3$\times$10$^{-9}$.  

Over 15 years, the SB profile of the disk may have varied, especially if there were any systematic changes in the illumination due to self-shadowing from the inner part of the disk or from the central star itself. We investigated whether there were any changes between the 2000 and 2015/2016 epochs (left panel of Figure \ref{fig:analysis}).  The disk SB does not appear to vary by more than 5-10\% between epochs from the inner working angle to the outer edge of the disk.

In addition to variability, we looked for radially coincident features in the TW Hya disk scattered light emission and the ALMA brightness temperature \citep[right panel of Figure \ref{fig:analysis} ][]{rapson15,andrews16}.  The breaks seen in SB are coincident with the 27 AU gap reported in the ALMA data, which is just outside the sub-mm transition between the optically thick inner disk and the optically thin outer disk \citep{nomura16,andrews16}.  The gap might be caused by a planet, especially since there is a lack of mm-sized dust observed at 22~AU \citep{tsukagoshi16}.  Alternatively, freezing-out and grain-growth can cause the apparent gap.  As the opacity to stellar irradiation decreases, this causes shadowing at the disk surface and cooling of the disk locally \citep[i.e.,][]{debes16,andrews16}.  Since stellar irradiation is the primary heat source in the disk at this distance, the reduction in temperature can propagate down to the midplane of the disk, especially if the disk is optically thin \citep{jangcondell12}.  Since the submillimeter continuum observations are sensitive to both disk temperature and density, such a temperature reduction would lower the local surface brightness.  The gap is likely to be narrower and deeper in the case of a density decrement as opposed to a pure temperature decrease, but fitting an exact model is beyond the scope of this paper.  Apart from the region near the 24~AU gap, there are no clear correlations between the STIS SB and the sub-mm emission.  

\subsection{Azimuthal Asymmetry Changes}
\label{sec:azimuth}
The TW Hya disk has a sinusoidal azimuthal asymmetry at certain radii \citep{Roberge:2005}, with a peak position angle that does not match the position angle of the semi-major disk axis of 155$^\circ$ \citep{andrews16,rosenfeld12}.  The asymmetry was further observed at larger radii in the same STIS dataset and also detected in an earlier NICMOS F110W image \citep{debes13}.  The origin of the SB asymmetry is debated and has been attributed to inclination effects for flared disks along the disk minor axis \citep{debes13}, the forward scattering properties of the disk \citep{Roberge:2005}, or the presence of a warp in the disk that modified the azimuthal surface brightness profile \citep{Roberge:2005,rosenfeld12}. 

If any inner disk structure blocks the light of the star, this can cause shadowing of the outer disk that modifies the azimuthal SB profile of a disk. An inner warp may also be a signature of a planet, since flaring and temperature variations in the disk are expected at the edges of planet gaps \citep{jangcondell12}. In this Section, we demonstrate that the position angle of the SB asymmetry peak in the TW~Hya disk is not stationary, but moves with time at disk radii $>$ 50~AU, placing constraints on the origin of the azimuthal scattered light asymmetry..  Further, we show that interior to 50~AU, there is a change in the nature of the SB azimuthal asymmetry. 

\begin{figure}
\plotone{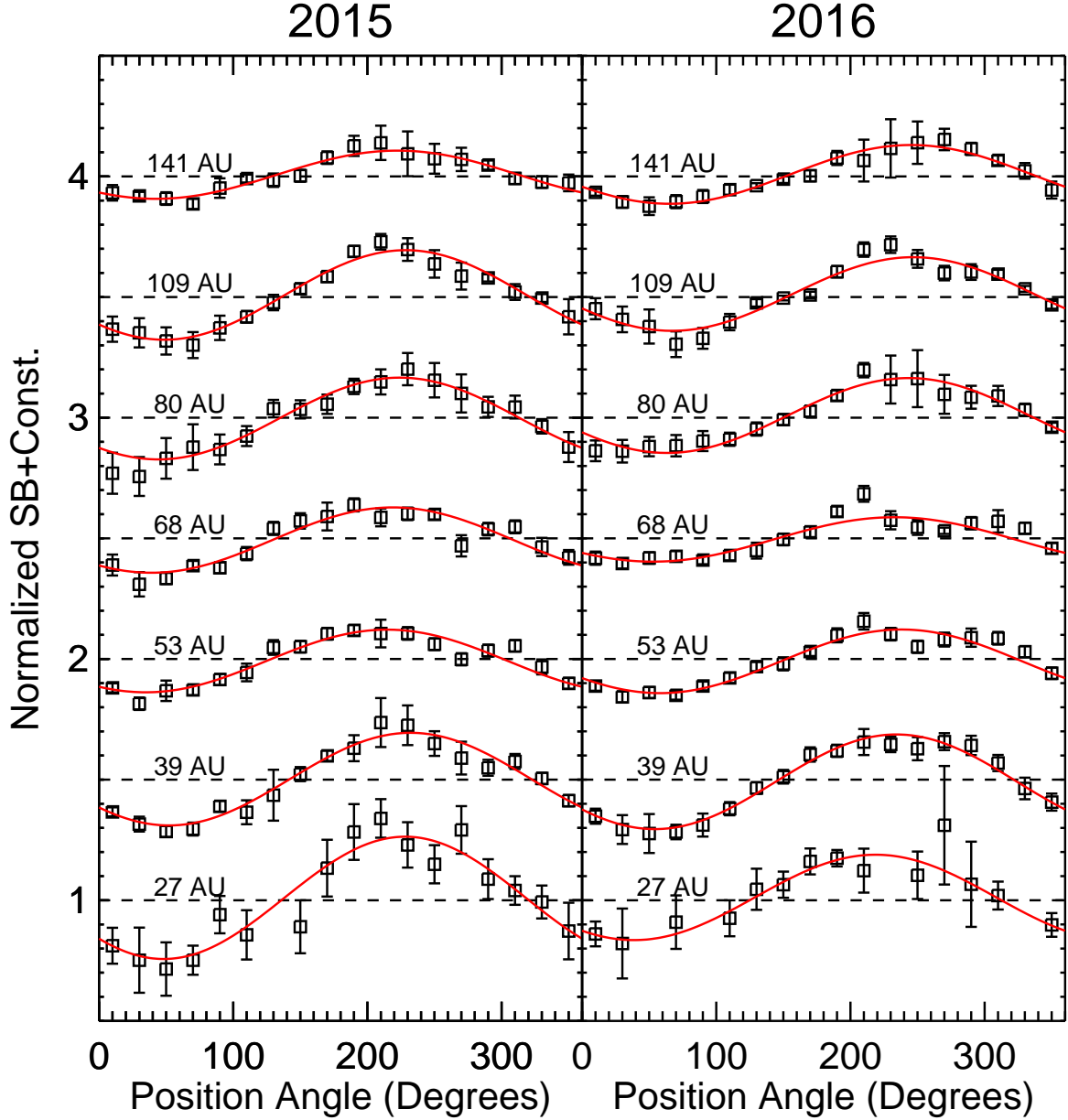}
\caption{\label{fig:asymmetries} Measured azimuthal asymmetries as a function of radius in the STIS data from epochs 2015 and 2016. We chose radii that matched seven regions of the disk centered at 0.46\arcsec\ (27~AU), 0.66\arcsec\ (39~AU), 0.89\arcsec\ (53~AU), 1.14\arcsec\ (68~AU), 1.47\arcsec\ (88~AU),1.83\arcsec\ (109~AU), and 2.36\arcsec\ (141~AU), with widths of 0.2\arcsec\ (12~AU), 0.2\arcsec\ (12~AU), 0.25\arcsec\ (15~AU), 0.25\arcsec\ (15~AU), 0.41\arcsec\ (24~AU), 0.30\arcsec\ (18~AU), and 0.56\arcsec\ (33~AU) respectively. Red curves show the best fitting cosine functions as described in the text and in Table \ref{tab:fits}.}
\end{figure}

To measure the properties of the asymmetries, we divided each azimuthal SB profile for a given radius by the mean SB at that radius to allow for unbiased radial averaging.  We chose radii that matched seven regions of the disk centered at 0.46\arcsec\ (27~AU), 0.66\arcsec\ (39~AU), 0.89\arcsec\ (53~AU), 1.14\arcsec\ (68~AU), 1.47\arcsec\ (88~AU), 1.83\arcsec\ (109~AU), and 2.36\arcsec\ (141~AU), with widths of 0.2\arcsec\ (12~AU), 0.2\arcsec\ (12~AU), 0.25\arcsec\ (15~AU), 0.25\arcsec\ (15~AU), 0.41\arcsec\ (24~AU), 0.30\arcsec\ (18~AU), and 0.56\arcsec\ (33~AU) respectively.  These were chosen to roughly correspond to regions within and near the two gaps seen in scattered light at 22 and 88~AU.
Figure \ref{fig:asymmetries} shows a comparison between our 2015 and 2016 data at the above distances, where azimuthal asymmetries are clearly detected in the data at all radii.  We first tested each radius to see if it had a significant departure from a constant profile by calculating its $\chi_\nu^2$ value. If the $\chi_\nu^2$ value of the azimuthal profile had a less than a 10$^{-3}$ probability of being consistent with a flat profile, we fit a cosine function of the form:
\begin{equation}
F(\theta) = A \cos \left[2 \pi (\theta-\theta_{\rm o})\right]+B
\end{equation}
where $\theta$ is the position angle as measured East through North on the sky, $A$ is the amplitude of the asymmetry, $\theta_{\rm o}$ is the position angle of the asymmetry peak, and $B$ is the constant offset to the curve.

 In all cases the resulting fit had an improved $\chi_\nu^2$ value that was close to 1. Table \ref{tab:fits} show our best-fit cosine parameters for each epoch and radius measured.

\begin{figure}
\plotone{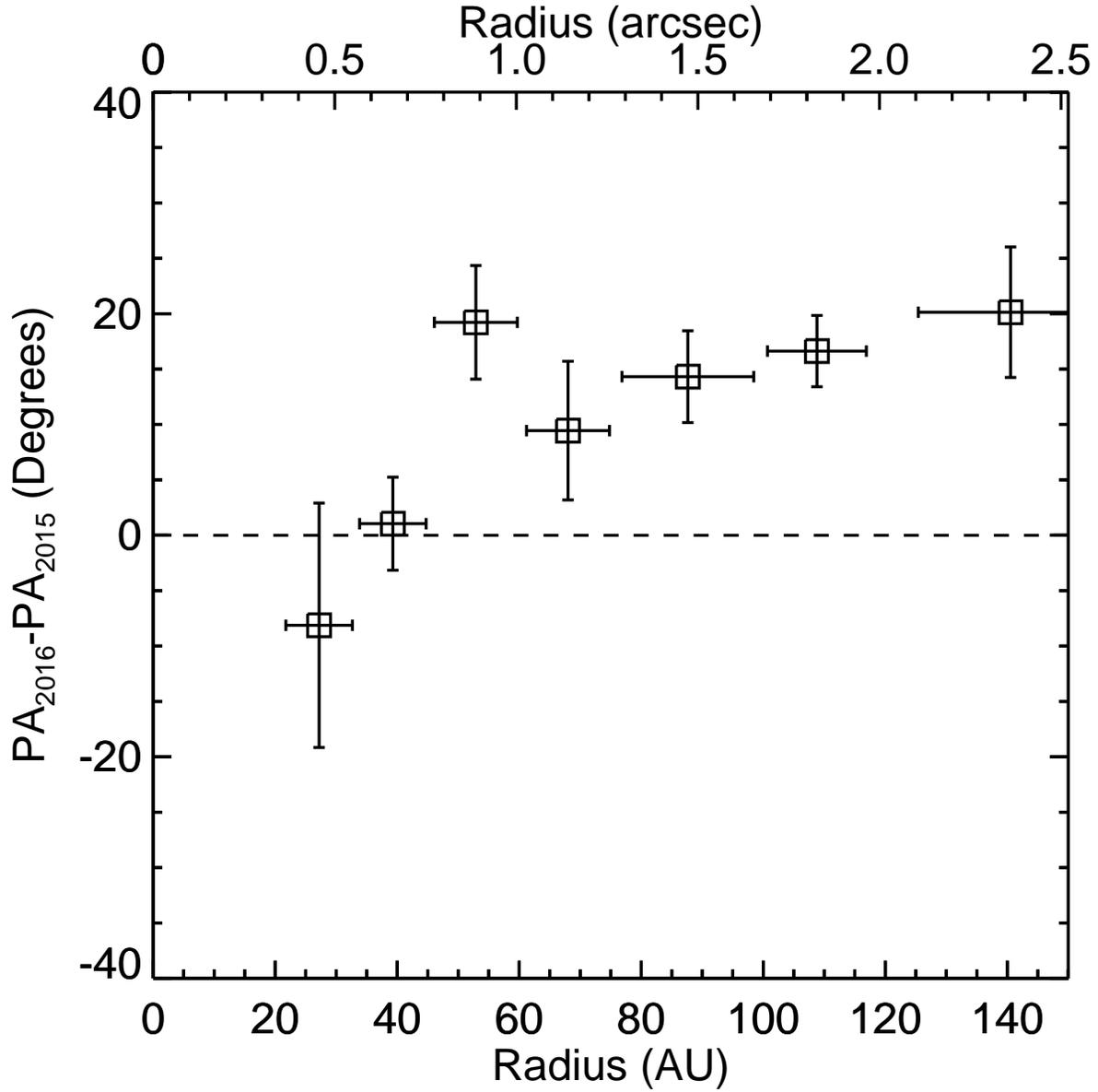}
\caption{\label{fig:diffs} Measured difference in the asymmetry peak PA as a function of radius in the TW Hya disk.  Beyond 50~AU, a clear difference in the peak position indicates that the asymmetry moves with an angular motion of 17$\pm$4$^\circ$ in a 10 month period, implying an instantaneous angular velocity of 20$\pm$5$^\circ$~yr$^{-1}$.}
\end{figure}

Figure \ref{fig:diffs} shows that between 2015 and 2016, the position angle of the azimuthal brightness peak shifted by a significant amount, on average 17$^\circ\pm4^\circ$, beyond 50~AU. Interior to 50~AU, there is no evidence of a change in the position angle of the asymmetry peak. Since the two epochs are roughly separated by 10 months, this translates to an instantaneous angular velocity of 20$^\circ\pm5^\circ$~yr$^{-1}$.

To further determine whether the azimuthal asymmetry was variable, we constructed similar SB asymmetry profiles from the 2000 STIS image, the 1998 F110W/F160W NICMOS images, the 2004 POL*L NICMOS total intensity (Stokes I) image, and the 2005 F171M/F180M/F222M NICMOS images.  We analyzed the NICMOS images in a similar fashion to the STIS data. 

For each filter we determined whether an asymmetry was significantly detected by calculating $\chi_\nu^2$ values as described for the two STIS epochs in 2015 and 2016.  With the exception of the F160W image, we found significant asymmetries for at least one radius at every filter wavelength, implying that the asymmetry is present from the optical through the near-IR.  The F160W images show no appreciable asymmetry, and we conducted tests that show the systematic noise within the image prevents the secure detection any similar asymmetry seen at other wavelengths. 

We also investigated whether a systematic PSF residual feature could masquerade as a moving asymmetry by measuring the PA asymmetry in  single spacecraft orientations.  For all of the STIS data, the asymmetry is fixed in sky orientation at a given epoch; i.e., the orientation of the asymmetry is not correlated with changing spacecraft orientation.  The two F110W spacecraft orientations were only separated by 7$^\circ$, but the average of the position angles are consistent with the asymmetry being fixed on the sky. For the NICMOS medium band and POL data, the dependence (or lack thereof) on spacecraft roll is less well determined, due to lower signal to noise.  If there were systematic residual features in the medium band and POL data, they would have to reside at roughly 100$^\circ$ in the medium band data and 140$^\circ$ in the POL data to mimic the behavior of the asymmetry's motion.

\begin{figure}
\plottwo{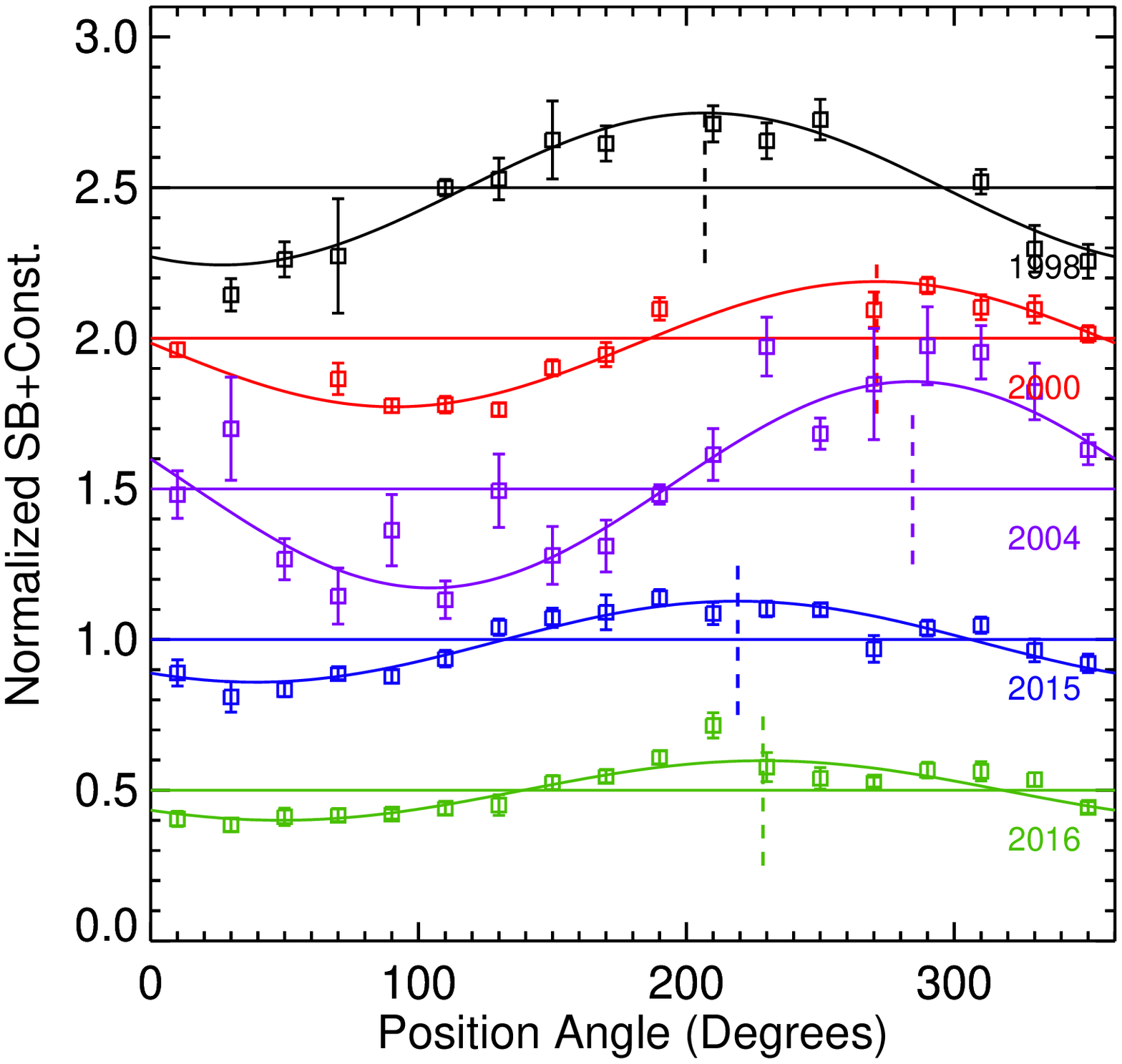}{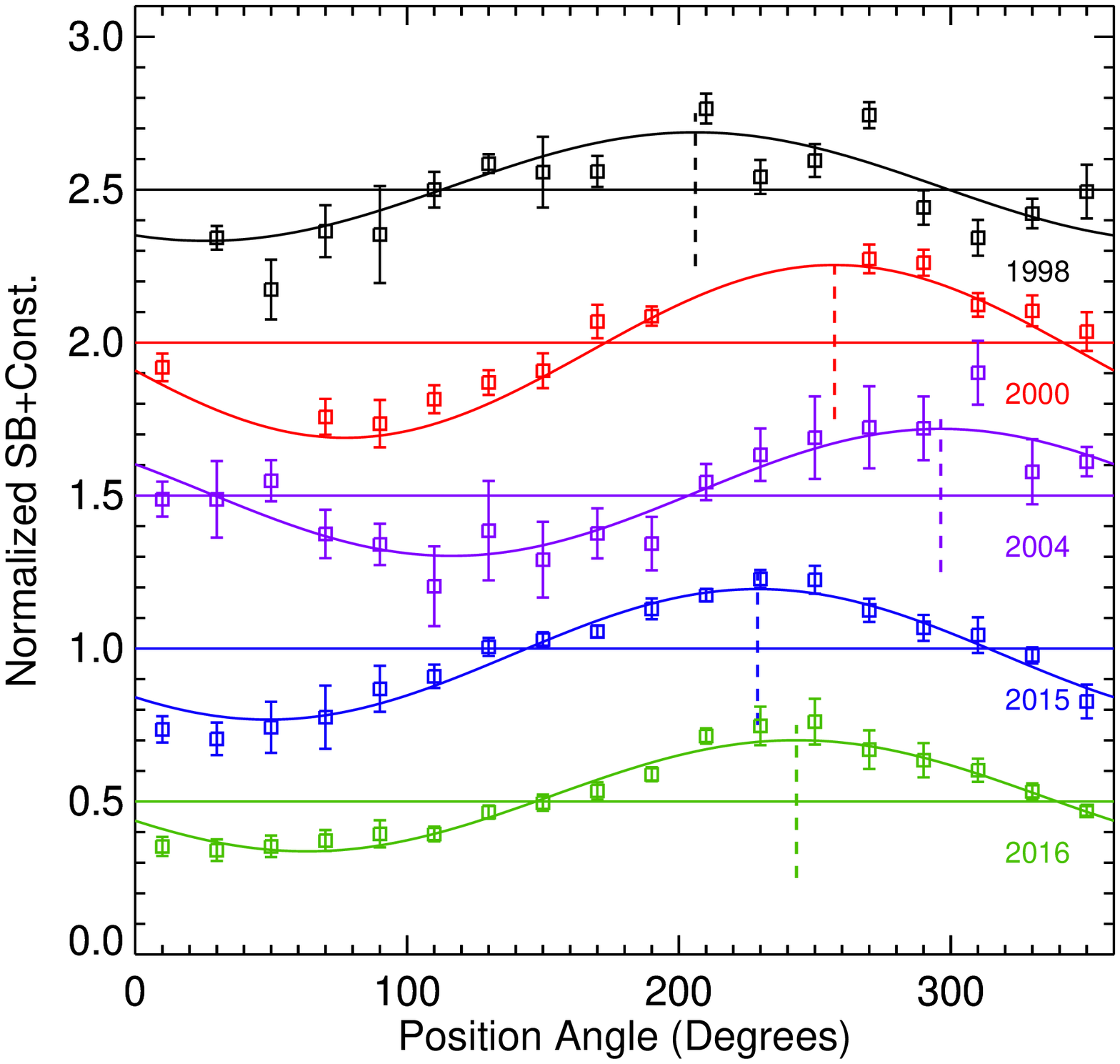}
\caption{\label{fig:az} The azimuthal asymmetry measured in the TW Hya disk at 88~AU (1.48\arcsec; left) and 109~AU (1.84\arcsec; right) from the star and averaged over radii with a widths of 0.41\arcsec\ (24~AU) and 0.30\arcsec\ (18~AU) between 1998 (NICMOS F110W), 2000 (STIS), 2004 (NICMOS POL filters), 2015 (STIS), and 2016 (STIS).  Dashed lines show asymmetry peak position angle and the solid lines are sinusoidal fits to the data.  The asymmetry is qualitatively similiar at the other distances.}
\end{figure}

\begin{figure}
\plotone{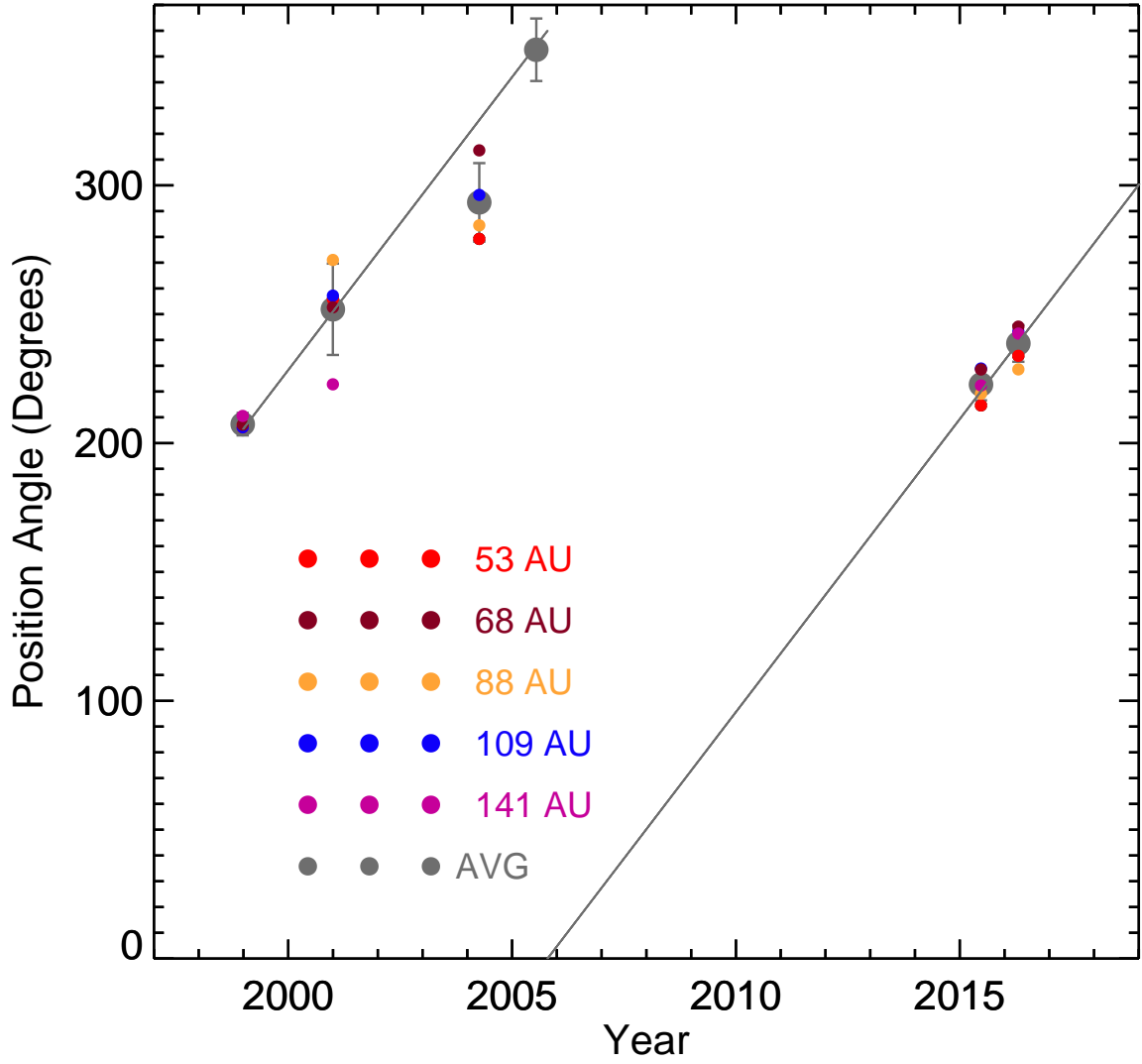}
\caption{\label{fig:vel} Plot of measured position angle vs. time for 53~AU (red), 68~AU (purple), 88~AU (orange), 109~AU (blue), and 141~AU (black).  Solid lines show the best fit position angle as a function of time, assuming a constant angular velocity of 22.7$^\circ$~yr$^{-1}$, or a period of 15.9~yr.  }
\end{figure}

Figure \ref{fig:az} shows the azimuthal profiles for five of the epochs with significantly detected asymmetries at 88~AU and 109~AU. The position angle of the peak in the azimuthal surface brightness profile clearly moves with time.  Sampling of the asymmetry is non-existent between 2005 and 2015, but the motion is consistent with constant rotation of the peak, as would be expected for circular orbital motion of a feature.  In Figure \ref{fig:vel} we plot the position angle of the asymmetry peak as a function of time at 53, 68, 88, 109, and 141~AU.  If we take an average of the position angles over radius and assume the scatter in the data represents the true uncertainty in the position angle measurement, a linear lest squares fit to the data results in a rate of 22.7$^\circ\pm0.4^\circ~$yr$^{-1}$ in a counter-clockwise direction, consistent with periods of 15.9$\pm0.3$~yr.  The measured velocities greatly exceed the local Keplerian velocity at the radii sampled (53-141~AU; expected angular velocities of $\sim$0.8$^\circ$-0.2$^\circ$~yr$^{-1}$), but would be consistent for an object orbiting at 5.6~AU.  

The recent detection of asymmetries in SPHERE data look qualitatively different from our total intensity images \citep{vanboekel}. It is difficult to do a direct comparison to these results since polarized intensity profiles may have different geometrical effects than those of total intensity images and may also suffer from different systematic uncertainties from PSF subtraction.  For example, the 2015 data from SPHERE was taken within a few months of our STIS data.  While the total intensity asymmetry does not appear to significantly change amplitude or shape with wavelength according to our data, the SPHERE polarized intensity data in the near-IR show a series of smaller peaks at roughly 130$^\circ$ and 220$^\circ$, with a decrement of flux at 300$^\circ$.  {\bf The azimuthal surface brightness profile in both polarized intensity and total intensity images can be modified by uncertainties in the stellar position relative to the reference PSF, and may explain these differences if they are not due to physical effects. We estimate that our uncertainty in centering ($\sim$0.1 pixels or 5 and 7~mas for STIS and NICMOS respectively) are small enough to only cause uncertainties in position angles of 5$^\circ$. Unfortunately, \citet{vanboekel} do not explicitly mention their estimated centering uncertainties for a direct comparison.}

\section{Interpretations of the Variable Azimuthal Asymmetry}
\label{sec:physical}

The angular velocity of the asymmetry is larger than the expected Keplerian velocity by at least an order of magnitude, and shows a constant velocity across a large swath of radii between 50-141~AU, yet no motion interior to 50~AU. A precessing warped inner disk that is at a higher inclination than the outer disk and is shadowing parts of the outer disk could explain the observed behavior in  TW~Hya. A warped inner disk at $\sim$5~AU was invoked to explain anomalous CO line emission in the disk, as well as the azimuthal asymmetry seen in the disk SB \citep{rosenfeld12}. The recent detection of a gap at $\sim$7~AU in scattered light is close to this location \citep{vanboekel}.  There has also been a tentative detection of a point source at ~6~AU with non-redundant aperture masking of the star \citep{willson16}. The period of the asymmetry motion, 15.9~yr, is comparable to the expected period from a circular orbit around TW~Hya at 5.6~AU.

We consider whether an inclined inner disk that shadows the outer part of the disk while precessing due to a planetary companion is a plausible explanation for what we observe. The inclination of the inner disk may not be significantly different than that of the outer disk, contrary to what is seen in HD 142725 \citep{marino15}, which only shows sharp shadows along the line of nodes for the inner, highly inclined disk.  Moderately inclined inner disks that were sufficiently flared or puffy could result in an asymmetry that roughly covers half of the outer disk, corresponding to where one lobe of the inner disk shadows the flared surface of the outer disk. This is what we see for TW~Hya.

The ALMA 870 $\mu m$ continuum emission image of TW Hya \citep{andrews16} provides some possible constraints on interpreting the location of the shadowing structure. Since we need to invoke an inclined inner disk, we note the most likely location is coincident with the unresolved emission detected around TW~Hya interior to 1~AU.  This inner disk has also been inferred to exist in interferometric datasets in the near-IR \citep{akeson11}. There are no obvious warps or departures from azimuthal symmetry in the sub-mm beyond 1~AU, and the interior of the system is preferable for shadowing a large range of grazing angles in the outer disk. 

To fully determine the plausibility of such an inclined disk and its impacts on the outer disk, detailed hydrodynamical modeling and radiative transfer would need to be done to self-consistently include the features observed in thermal emission and scattered light.  While such modeling is beyond the scope of this paper, we can assume that the outer disk is well approximated from the best-fit model of TW Hya presented in \citet{debes13}, and that the inner inclined disk has an opening angle that is constrained by the extent of the shadow asymmetry. The inner disk could be torqued by a companion on an inclined orbit and it will precess due to the interaction between the companion's orbit, the spin of the star, and the disk \citep[e.g][]{lai14}. Based on the best fit model of TW~Hya's disk in \citet{debes13}, the height $H$ of the $\tau=2/3$ surface above the midplane for the disk in the STIS bandpass is 12~AU at a radius $R$=53~AU and $H$=34~AU at $R$=141~AU, the rough extent to which we recover the shadow asymmetry in all three STIS epochs.  We determine the range of inclinations relative to the disk midplane that cause the shadow by calculating $\delta = \sin^{-1}\left(\frac{H}{R}\right)$. If we assume that the inner disk midplane is in the middle of the angles probed, the inferred inclined disk must span from $\sim$13$^\circ$ to $\sim$14$^\circ$. We note that this is compared to the plane of the outer disk--in reality the outer disk is inclined from our line of sight by 7$^\circ$ \citep{qi04}.

The lack of rotation in the asymmetry interior to 53 AU could be due to several factors: 1) the opening angle of the inner disk is larger and the difference in inner vs. outer disk inclinations is smaller, such that the inner disk casts shadows over all azimuthal angles; 2) the outer disk is fully illuminated by the star and thus there exists a puffed inner rim at $\sim$2~AU, shadowing the disk from 2-53~AU; or 3) the inner edges of one of the other gaps (13 or 22~AU) are shadowing the region between 27-53~AU. In any case, the asymmetry present at these radii has a peak at a position angle of 232$^\circ$, close to the expected minor axis of the disk. Given this alignment, the origin of the (stationary) asymmetry interior to ~50 AU presumably lies in a combination of inclination effects and the proximity to the gap at ~25 AU discussed previously\citep{a15,rapson15,debes16}.

The period of rotation of the asymmetry, if caused by precession of an inclined inner disk due to a planetary companion, is dictated by the mass of the companion, the extent of the perturbed disk, and the orbit of the companion.  Given the preceding model for the rotating outer disk shadow pattern that invokes an inclined inner disk, we can place tentative constraints on the mass of the putative perturbing body. The precession period is given by \citep{lai14}:

\begin{equation}
\label{eq:prec}
P=\frac{2.04\times10^{5}}{\mu_{\rm c} \cos{i_{\rm c}}} \left(\frac{M_\star}{M_\odot}\right)^{-\frac{1}{2}}\left(\frac{r_{\rm disk}}{1 {\rm AU}}\right)^{-\frac{3}{2}}\left(\frac{a_{\rm c}}{300~{\rm AU}}\right)^3~{\rm yr}
\end{equation}

where $\mu_{\rm c}$ is the mass ratio of the perturber to the central star, i$_c$ is the inclination of the binary relative to the inner disk, $M_\star$ is the mass of the central star (0.69 M$_{\odot}$), $r_{disk}$ is the extent of the inner disk, and $a_{\rm c}$ is the semi-major axis of the perturber. Solving for the perturber mass with $i_c$=14$^\circ$, $P$=15.9~yr, $r_{\rm disk}$=1~AU, and $a_c$=1.1~AU results in a mass of 1.2~M$_{\rm Jup}$.  If there were a companion at $a_c$=7~AU, with $r_{\rm disk}$=6~AU, the mass of the companion would need to be 19~M$_{\rm Jup}$ to create a 15.9~yr precession period in the inner disk. The high mass needed at this distance suggests that the location of the inclined disk is closer to 1~AU than at this location if disk precession is the cause of the rotating shadow.

The radial velocity semi-amplitude and period for such a planetary companion at 1.1~AU would be $\approx$10~m s$^{-1}$ and $\approx$1.4~yr, respectively which is consistent with the non-detection of any radial velocity trend in TW~Hya at 6 m~s$^{-1}$ precision over a six-day period \citet{huelamo08,figueira10}. The proposed structure of the disk here would also need to be reconciled with existing interferometric data in the near-IR, mid-IR, and at cm wavelengths, as well as with the reported gap in scattered light reported by \citet{vanboekel}.

The sparse temporal coverage of the extant HST scattered light imaging of the TW Hya disk leaves room for other potential behavior for the asymmetry that may mimic rotation at constant angular velocity. It is possible that the inferred shadow pattern rotation is due either to a much shorter-period phenomenon or to some stochastic variation. However, we have inspected the individual STIS images obtained in 2000 and the individual STIS images of 2015 and 2016, and we find no significant variations on such (month to year) timescales. If we assume that we would have been able to detect azimuthal brightness pattern motion in these images at the level of 5-10$^\circ$ per epoch, then the lack of detectable changes within individual epochs appears to rule out the subset of stochastic or very rapid variations that might occur on hourly to monthly timescales.  

We can also rule out the possibility that the observed linear trend in position angle with time is the result of observations at a completely random set of position angles.  We constructed a Monte Carlo model of random position angles as a function of time at the six epochs we observed TW Hya. We performed 10$^6$ trials of randomly selecting position angles to see what fraction could still be fit with a linear slope where the $chi^2$ value implied a probability of a good fit to better than a probability of 5\%.  From these 10$^6$ trials, we find a 4$\times$10$^{-5}$ probability that the slope is a random occurrence.

While it is possible that the asymmetry is only due to a shadow, inclination effects can also modify the inferred motion of a shadow asymmetry.  As mentioned in Section \ref{sec:azimuth}, the inclination of a flared disk can result in a weak azimuthal asymmetry that can become more pronounced as the inclination departs from face-on or in the presence of forward scattering dust grains. The combination of the shadow and the inclination asymmetries for small inclinations can be approximated analytically:

\begin{equation}
\label{eq:beat}
SB_{\rm disk} \approx SB_{\rm o}\left[1+A_{1}\sin{\theta}+A_{2}\sin{\left(\theta+\alpha\right)}\right]
\end{equation}
where $A_1$ is the amplitude of the inclination asymmetry, $A_2$ is the shadow asymmetry amplitude, $\theta$ is the sky position angle, and $\alpha$ is the position angle as a function of time for the shadow. This simplifies to:
\begin{equation}
SB_{\rm disk}\approx SB_{\rm o}\left[1+C \sin{\left(\theta+\phi\right)}\right]
\end{equation}
where the amplitude of the asymmetry is:
\begin{equation}
C = \sqrt{A_{1}^2+A_{2}^2+2A_{1}A_{2}\cos{\alpha}}
\end{equation}
and the observed position angle of the asymmetry is given by:
\begin{equation}
\label{pa}
\phi=\tan^{-1}{\frac{A_{2}\sin{\alpha}}{A_{1}+A_{2}\cos{\alpha}}}
\end{equation}

If $A_1> A_2$, then one should expect the observed asymmetry to change in amplitude and even disappear at certain epochs, while librating about the observed position angle of the minor axis of the disk.  Conversely, if  $A_{1}<A_2$, the asymmetry will be constant with time and circulate at the angular velocity of the shadow. In the case of TW~Hya, given the low expected inclination of the outer disk, $A_1$ is a factor of $\sim$3 smaller than the amplitudes actually measured. No obvious change in amplitude is observed epoch-to-epoch between the STIS images. There are also no significant deviations from a purely linear trend in the measured position angle of the asymmetry, as would be expected for a shadow amplitude that is slightly larger than the inclination asymmetry.

\section{Conclusions}
\label{sec:conc}
With STIS+NICMOS, we can probe the changing illumination of the protoplanetary disk of TW~Hya with unprecedented detail on timescales of 17~yr.  The new BAR5 occulter on STIS probes stellocentric distances comparable to the orbital radius of Saturn.  While the SB profile of the disk has remained constant over 15~yr, the azimuthal asymmetry previously seen in the disk is variable in position angle, and best described by the orbital motion of a shadowing structure in the inner parts of the disk.  

These observations are part of a growing list of scattered light variability observed in protoplanetary transition disks with rings and gaps.  For example, PDS 66 showed variability in its SB in the optical \citep{schneider14} on monthly timescales and hints at azimuthal movement of features with GPI polarized intensity data \citep{wolff16}.  The transition disk J160421.7-213028, observed with SPHERE/ZIMPOL in the visible and with HiCIAO in polarized H band intensity, also showed a localized dip in the SB of its disk that appeared to move by $\sim$12$^\circ$~yr$^{-1}$; however, the angular velocity of the shadow pattern is indeterminate, given that data were obtained at only two epochs \citep{pinilla15,mayama12}.  Additional claims of variable shadowing have been put forth for HD 163296 \citep{wisniewski08}, but again the epoch-to-epoch variation was hard to characterize due to a dearth of coverage.  

The combination of the broad azimuthal extent of the asymmetry, its rotational motion, and the range of radii over which it appears is suggestive of the presence of an anomalously inclined structure in the disk interior to 5 AU that shadows the outer regions of the disk. Given the inferred rotational timescale of the azimuthal asymmetry, we propose that it is the result of shadowing by an inclined disk of radius <1 AU that is precessing with an orbital period of 15.9 yr. Adopting reasonable assumptions concerning the inclination of the disk and its precession due to an external perturber at 1.1 AU, we estimate the mass of a putative perturbing body to be $\sim$1~M$_{\rm Jup}$. Such a mass is small enough to have thus far escaped detection by previous campaigns to measure radial velocity variations in TW Hya.

Our results for TW Hya demonstrate that the detailed structure of protoplanetary disks within 20 AU impact the scattered light emission observed in the outer disk.  Time domain high contrast scattered-light imaging with high signal-to-noise and small inner working angles, such as the observations of TW Hya with STIS, GPI, and SPHERE, provides an approach complementary to that of future observations of disks that will be obtained with ALMA and the {\em James Webb Space Telescope}.

\acknowledgements
Support for this work was provided by NASA through grants HST-GO-13753, and  from the Space Telescope Science Institute, which is operated by AURA, Inc., under NASA contract NAS 5-26555. J.H.K.'s research on protoplanetary disks orbiting nearby young stars is supported by NASA Exoplanets Program grant NNX16AB43G to the Rochester Institute of Technology. This research has made use of the VizieR catalogue access tool and the SIMBAD database, CDS, Strasbourg, France. The original description of the VizieR service was published in A\&AS 143, 23. This work has made use of data from the European Space Agency (ESA)
mission {\it Gaia} (\url{http://www.cosmos.esa.int/gaia}), processed by
the {\it Gaia} Data Processing and Analysis Consortium (DPAC,
\url{http://www.cosmos.esa.int/web/gaia/dpac/consortium}). Funding
for the DPAC has been provided by national institutions, in particular
the institutions participating in the {\it Gaia} Multilateral Agreement.
This manuscript benefited from discussions with Steven Lubow.


\begin{deluxetable}{cccccccc}
\tablecolumns{8} 
\tablewidth{0pt} 
\tabletypesize{\scriptsize}
\tablecaption{\label{tab:obslist} New and archival HST observations of TW Hya}
\tablehead{\colhead{MJD} & \colhead{UTC Date} & \colhead{Instrument} & \colhead{Filter} & \colhead{Ref. Star} & \colhead{F$_{\star}$/F$_{\rm ref}$} & \colhead{T$_{\rm exp}$ (s)} & \colhead{Orientation}}
\startdata
51041 & 16-August-1998 & NICMOS & F160W & GL 879 & 0.031 & 1213  & -72.7,-65.7\\
51142 & 25-November-1998 & NICMOS & F110W & $\tau^1$ Eridani & 0.099 & 1213  & 52.5, 59.5\\
51851 & 03-November-2000 & STIS & 50CCD & HD 85512 & 0.055 & 2164 & -139.9 \\
51905 & 27-December-2000 & STIS & 50CCD & HD 85512 & 0.057 & 2164 & -106.9 \\
53037 & 02-February-2004 & NICMOS & POL*L\tablenotemark{a} & BD$+$32$\degr$3739 & 3.5\tablenotemark{b} &  863 & 88.6,118.5 \\
53108 & 04-April-2004 & NICMOS & POL*L & BD$+$32$\degr$3739 & 3.4 & 863 & 182.7,212.6 \\ 
53538 & 17-June-2005 & NICMOS & ...\tablenotemark{c} & CD-43 2742 & ...\tablenotemark{d} &  3712  & -125.4, -97.4 \\
57166 & 24-May-2015 & STIS & 50CCD & HD 85512 & 0.051 & 5661  & 49.0, 64.0, 79.0\\
57471 & 24-March-2016 & STIS & 50CCD & HD 85512 & 0.055 & 6381 & -31.9,-16.9, 3.8\\
\enddata
\tablenotetext{a}{The POL filters are POL0L, POL120L, and POL240L}
\tablenotetext{b}{The scaling factors for the POL filters were averaged over all spacecraft orientations and filters}
\tablenotetext{c}{F171M, F180M, F222M}
\tablenotetext{d}{See \citet{debes13} for details of the scaling for these filters} 
\end{deluxetable}

\begin{deluxetable}{cccccccc}
\tablecolumns{8} 
\tablewidth{0pt} 
\tablecaption{\label{tab:fits} Summary of Best-Fit Asymmetry Parameters}
\tablehead{\colhead{MJD} & \colhead{Filter} & \colhead{R (AU)} & \colhead{Amplitude}  & \colhead{PA ($\degr$)} & \colhead{offset} & \colhead{$\chi^2_{\nu,fit}$} & \colhead{$\nu$}}
\small
\startdata
51142 & F110W &  38 &  0.35 $\pm$  0.12 &  192 $\pm$   4 &   1.15 $\pm$   0.08 &   0.62 &   5 \\ 
51142 & F110W &  51 &  0.24 $\pm$  0.03 &  206 $\pm$   4 &   0.99 $\pm$   0.02 &   1.87 &   9 \\ 
51142 & F110W &  67 &  0.25 $\pm$  0.02 &  207 $\pm$   4 &   1.00 $\pm$   0.02 &   1.33 &  10 \\ 
51142 & F110W &  87 &  0.18 $\pm$  0.02 &  206 $\pm$   5 &   1.01 $\pm$   0.01 &   3.02 &  13 \\ 
51142 & F110W & 112 &  0.20 $\pm$  0.02 &  207 $\pm$   3 &   1.02 $\pm$   0.01 &   5.73 &  15 \\ 
51142 & F110W & 141 &  0.21 $\pm$  0.01 &  211 $\pm$   3 &   1.03 $\pm$   0.01 &  13.02 &  15 \\ 
51878 & STIS &  39 &  0.20 $\pm$  0.02 &  243 $\pm$   3 &   1.02 $\pm$   0.01 &   3.04 &   7 \\ 
51878 & STIS &  53 &  0.20 $\pm$  0.02 &  256 $\pm$   3 &   0.97 $\pm$   0.01 &   0.77 &   7 \\ 
51878 & STIS &  68 &  0.21 $\pm$  0.01 &  271 $\pm$   3 &   0.98 $\pm$   0.01 &   1.91 &  10 \\ 
51878 & STIS &  88 &  0.28 $\pm$  0.01 &  257 $\pm$   2 &   0.97 $\pm$   0.01 &   2.35 &  15 \\ 
51878 & STIS & 109 &  0.21 $\pm$  0.02 &  253 $\pm$   3 &   1.02 $\pm$   0.01 &   0.48 &  15 \\ 
51878 & STIS & 141 &  0.07 $\pm$  0.01 &  223 $\pm$   7 &   1.00 $\pm$   0.01 &   1.84 &  15 \\ 
53072 & POL* &  38 &  0.14 $\pm$  0.02 &  260 $\pm$   7 &   1.00 $\pm$   0.01 &   1.55 &  15 \\ 
53072 & POL* &  51 &  0.20 $\pm$  0.02 &  279 $\pm$   5 &   1.03 $\pm$   0.01 &   0.98 &  15 \\ 
53072 & POL* &  67 &  0.34 $\pm$  0.03 &  284 $\pm$   3 &   1.01 $\pm$   0.02 &   1.95 &  15 \\ 
53072 & POL* &  87 &  0.21 $\pm$  0.03 &  296 $\pm$   6 &   1.01 $\pm$   0.02 &   0.85 &  15 \\ 
53072 & POL* & 112 &  0.13 $\pm$  0.02 &  314 $\pm$  11 &   1.02 $\pm$   0.02 &   1.78 &  15 \\ 
53072 & POL* & 141 &  0.14 $\pm$  0.04 &   16 $\pm$  19 &   0.99 $\pm$   0.02 &   2.00 &  15 \\ 
53538 & Various &  51 &  0.09 $\pm$  0.02 &  353 $\pm$  12 &   1.01 $\pm$   0.01 &   2.72 &  15 \\ 
57166 & STIS &  27 &  0.25 $\pm$  0.03 &  228 $\pm$   7 &   1.01 $\pm$   0.02 &   0.71 &  14 \\ 
57166 & STIS &  39 &  0.19 $\pm$  0.01 &  232 $\pm$   2 &   1.00 $\pm$   0.01 &   0.89 &  15 \\ 
57166 & STIS &  53 &  0.13 $\pm$  0.01 &  215 $\pm$   3 &   0.99 $\pm$   0.01 &   1.82 &  15 \\ 
57166 & STIS &  68 &  0.13 $\pm$  0.01 &  219 $\pm$   4 &   0.99 $\pm$   0.01 &   1.53 &  15 \\ 
57166 & STIS &  88 &  0.21 $\pm$  0.01 &  229 $\pm$   3 &   0.98 $\pm$   0.01 &   1.33 &  15 \\ 
57166 & STIS & 109 &  0.18 $\pm$  0.01 &  229 $\pm$   2 &   1.01 $\pm$   0.01 &   0.78 &  15 \\ 
57166 & STIS & 141 &  0.10 $\pm$  0.01 &  222 $\pm$   4 &   1.01 $\pm$   0.01 &   0.65 &  15 \\ 
57471 & STIS &  27 &  0.18 $\pm$  0.03 &  220 $\pm$   8 &   1.02 $\pm$   0.02 &   0.47 &  14 \\ 
57471 & STIS &  39 &  0.20 $\pm$  0.01 &  233 $\pm$   2 &   0.99 $\pm$   0.01 &   0.52 &  15 \\ 
57471 & STIS &  53 &  0.14 $\pm$  0.01 &  234 $\pm$   3 &   0.99 $\pm$   0.01 &   1.07 &  15 \\ 
57471 & STIS &  68 &  0.10 $\pm$  0.01 &  229 $\pm$   4 &   1.00 $\pm$   0.01 &   2.21 &  15 \\ 
57471 & STIS &  88 &  0.18 $\pm$  0.01 &  243 $\pm$   2 &   1.02 $\pm$   0.01 &   1.05 &  15 \\ 
57471 & STIS & 109 &  0.17 $\pm$  0.01 &  245 $\pm$   1 &   1.01 $\pm$   0.01 &   2.43 &  15 \\ 
57471 & STIS & 141 &  0.12 $\pm$  0.01 &  242 $\pm$   3 &   1.01 $\pm$   0.01 &   0.76 &  15 \\ 
\enddata
\end{deluxetable}


\end{document}